\documentclass[sigconf]{acmart}

\usepackage{booktabs}	
\usepackage{diagbox}    
\usepackage{multirow}   
\usepackage{verbatim}	
\usepackage{array}    
\usepackage{balance} 

\usepackage{amsmath,amssymb,amsfonts,graphicx}
\usepackage{epstopdf}

\copyrightyear{2023}
\acmYear{2023}
\setcopyright{acmlicensed}\acmConference[WWW '23 Companion]{Companion Proceedings of the ACM Web Conference 2023}{April 30-May 4, 2023}{Austin, TX, USA}
\acmBooktitle{Companion Proceedings of the ACM Web Conference 2023 (WWW '23 Companion), April 30-May 4, 2023, Austin, TX, USA}
\acmPrice{15.00}
\acmDOI{10.1145/3543873.3587593}
\acmISBN{978-1-4503-9419-2/23/04}

\begin{document}

\title{The Human-Centric Metaverse: A Survey}



\author{Riyan Yang}
\authornote{Both authors contributed equally to this work.}
\affiliation{%
	\institution{Jinan University} 
	\city{Guangzhou}
	\country{China}
}
\email{ryyang66@gmail.com}

\author{Lin Li}
\authornotemark[1]
\affiliation{%
	\institution{Jinan University} 
	\city{Guangzhou}
	\country{China}
}
\email{linli.wh@gmail.com}

\author{Wensheng Gan}
\authornote{Also with Pazhou Lab, Guangzhou 510330, China}
\affiliation{%
	\institution{Jinan University} 
	\city{Guangzhou}
	\country{China}
}
\email{wsgan001@gmail.com}

\author{Zefeng Chen}
\affiliation{%
	\institution{Jinan University} 
	\city{Guangzhou}
	\country{China}
}
\email{czf1027@gmail.com}

\author{Zhenlian Qi}
\authornote{Corresponding author: qzlhit@gmail.com}
\affiliation{%
	\institution{Guangdong Eco-Engineering Polytechnic}
	\city{Guangzhou}
	\country{China}
}

\renewcommand{\shortauthors}{Yang \textit{et al.}}

\begin{abstract}
  In the era of the Web of Things, the Metaverse is expected to be the landing site for the next generation of the Internet, resulting in the increased popularity of related technologies and applications in recent years and gradually becoming the focus of Internet research. The Metaverse, as a link between the real and virtual worlds, can provide users with immersive experiences. As the concept of the Metaverse grows in popularity, many scholars and developers begin to focus on the Metaverse's ethics and core. This paper argues that the Metaverse should be centered on humans. That is, humans constitute the majority of the Metaverse. As a result, we begin this paper by introducing the Metaverse's origins, characteristics, related technologies, and the concept of the human-centric Metaverse (HCM). Second, we discuss the manifestation of human-centric in the Metaverse. Finally, we discuss some current issues in the construction of HCM. In this paper, we provide a detailed review of the applications of human-centric technologies in the Metaverse, as well as the relevant HCM application scenarios. We hope that this paper can provide researchers and developers with some directions and ideas for human-centric Metaverse construction.
\end{abstract}



\maketitle

\section{Introduction}

With the development of the Web of Things, the Metaverse \cite{sun2022metaverse,sun2022big,chen2022metaverse} is a new type of Internet application and social form. It integrates many new technologies, such as data science \cite{gan2017data,gan2019survey,wan2022fast}, artificial intelligence (AI) \cite{pham2022artificial,chen2022federated}, the Internet of Things \cite{mozumder2022roadmap,sun2023internet}, blockchain \cite{gadekallu2022blockchain}, cloud computing \cite{tang2022roadmap}, augmented reality (AR) \cite{azuma1997survey}, virtual reality (VR) \cite{schuemie2001research}, and digital twins \cite{far2022applying}, etc. It creates a virtual world that maps to the real world. In general, the Metaverse is regarded as a fully immersive, hyperspace, self-sustaining virtual space that integrates physical, human, and digital elements. By closely combining the virtual world with the real world, users can use digital avatars in the Metaverse to have a unique experience. Metaverse is recognized as an evolving paradigm of the next-generation Internet after the web, and its different participants are enriching its meaning in their own ways.

The Metaverse is immature and in its infancy. It involves many technologies in a wide range and takes time to realize them step by step. How to build a Metaverse nowadays becomes an important problem that needs to be solved \cite{laeeq2022metaverse}. In our opinion, the development and even survival of the Metaverse are human-centric \cite{mourtzis2022human}. In other words, the Metaverse always serves humans, and it will lose its meaning without the existence of human beings. For example, the construction of the Metaverse needs to take into account human experience \cite{george2021metaverse}; the operation of the Metaverse requires the organization and participation of human beings; although the Metaverse is a virtual world, it still needs to consider the various rights of human beings, and it needs to fight against crimes \cite{mourtzis2022human}, etc. Since the conception and ideas of the Metaverse come from human beings, the Metaverse needs to follow the principle of putting people first. In fact, the beginning of Metaverse opened up products and services that connect the virtual world and the real world, and this is about serving human beings. Metaverse is the product of the interconnection between the virtual world and the real world, and it is also a tool for human beings to explore the world and communicate with each other in the world. 

The construction and development of the Metaverse ought to follow the human-centric principle \cite{wang2022survey}, and the specific performance is as follows. Firstly, the Metaverse should be humanized and people-oriented. At its core, the Metaverse is humane and should be dedicated to helping others, not extending the interests of entrepreneurs and system designers. The development of the Metaverse should start with human experience and determine how the world should work \cite{mozumder2022overview}. There are companies whose short-term profit motives may conflict with this vision, which is unhealthy and unsustainable. Secondly, the operation of the Metaverse must be based on the group's standards. The owners of the Metaverse must be people instead of a corporation. The Metaverse advocates for people to be able to create rather than just consume \cite{ondrejka2004escaping}. For companies seeking long-term growth, this is a must and should be a guideline. Thirdly, accessibility and inclusion are imperatives in the Metaverse. Thirdly, the Metaverse should be a victory for diversity, equity, and inclusion (DEI) \cite{zallio2022designing}. The Metaverse enables anyone to participate in various activities in it, and not only those with status and money can enjoy it, which is a vision for the future Metaverse. In other words, people regardless of skin color, nationality, industry, status, or gender can freely discuss and communicate in the future Metaverse. If the future Metaverse can be human-centric, it will have considerable potential. As long as technologies are sufficient to support the functioning of the Metaverse, the Metaverse enables them to span distances in this virtual world \cite{njoku2023prospects}.

\textbf{Contributions}: We are convinced that the essence of the Metaverse is to create a human-centric future social ecology that integrates the virtual and real worlds, or to create a human-centric Metaverse (HCM). We believe that the human-centric Metaverse remains a powerful force for long-term positive changes now and in the future. This research focuses on the opportunities for HCM. The main contributions of this paper are as follows.

\begin{itemize}
    \item We discuss why we should be human-centric and introduce the concept of the human-centric Metaverse. We also introduce and discuss the characteristics and key technologies of the Metaverse. This is the first review for HCM.

    \item We summarize how the technology of the Metaverse embodies human-centricity and then introduce the key technologies of the HCM in detail. Furthermore, how the technologies could improve to make the Metaverse more human-centric is discussed in-depth.

    \item We detail the application scenario of the HCM in reality. Additionally, some open problems of the human-centric Metaverse are highlighted, as well as some opportunities.

    \item Finally, we give a conclusion to this paper and set out our future vision for HCM. 
\end{itemize}

\textbf{Organization}: The rest of the paper is organized as follows. In section \ref{sec:Metaverse}, we briefly describe the Metaverse's concept, characteristics, and technical support. Then, in Section \ref{sec:HCM}, we put forward the human-centric Metaverse, including why it should be human-centric and what is the human-centric Metaverse. In Section \ref{sec:Technologies}, we present the human-centric concept in the Metaverse technology. Section \ref{sec:Application} summarizes the application scenarios of the human-centric Metaverse in reality. Then, in Section \ref{sec:Problems}, we discuss the problems facing the construction of the human-centric Metaverse. Finally, we summarize the entire paper in Section \ref{sec:conclusion}.

\section{Metaverse}  \label{sec:Metaverse}

"Metaverse"\footnote{https://en.wikipedia.org/wiki/Metaverse} was first proposed by Neal Stephenson in the cyberpunk science fiction book \textit{Snow Crash} in 1992. In this novel, the Metaverse is a virtual cyberspace parallel to the real universe, where everyone can experience it immersively through their network avatars. Metaverse is composed of two roots, \textit{Meta} and \textit{verse} \cite{wang2022survey}. \textit{Meta} means transcendence in Latin, and the \textit{verse} is taken from the universe, which means beyond the present universe, that is, the future universe. At the moment, the Metaverse is the post-reality universe, a perpetual and persistent multiuser environment merging physical reality with digital virtuality \cite{mystakidis2022metaverse}, and it is the trend of Internet development in the future. The Metaverse is a new norm of network platform that integrates social networking, public services, intelligent manufacturing, medical health, education, and other functions \cite{gadekallu2022blockchain}. It is also a unity of information technology, a comprehensive technical form with blockchain, artificial intelligence, VR/AR technology, digital twins, and other technologies as the underlying architecture \cite{mozumder2022overview}. Its existence is not to separate from reality but to emphasize a new form of the Internet that is symbiotic between the virtual and the real and human-centric \cite{huggett2020virtually}.

\subsection{Characteristics of the Metaverse}

As a new Internet application, the Metaverse integrates various new-era technologies, combining multiple disciplines and industries. Specifically, the Metaverse shows unique characteristics from the following perspectives:

\textbf{\emph{1)  Immersion:}} Immersion is the key to breaking the real and virtual walls between the Metaverse and the real world. With the development of AR, VR, and other technologies, computer-generated virtual space can allow users to immerse themselves psychologically and emotionally. In the future, people will receive information through hearing, vision, touch, and perceive the world in the Metaverse through the third dimension of touch \cite{dionisio20133d, maccallum2019teacher, dincelli2022immersive}.

\textbf{\emph{2)  Hyper spatio-temporally:}} It is a term used for existence in both space and time \cite{sriram2022comprehensive}, which refers to breaking the limits of both time and space. However, the Metaverse is a virtual network world parallel to the real world \cite{ning2021meteverse}, where users can experience different lives in different time and space latitudes.

\textbf{\emph{3)  Sustainability:}} Numerous enterprises and developers jointly construct the Metaverse's common underlying standards and value norms \cite{wang2021research}. It has the characteristics of high liberalization and decentralization. The bankruptcy of individual enterprises or the withdrawal of creators will only affect the survival of part of the independent small world in the Metaverse. The construction and development of the platform will not be delayed and will continue to develop indefinitely in an open-source and transparent way.

\textbf{\emph{4)  Closed economic system:}} It indicates that the Metaverse maintains a closed economic loop and a consistent value system with a high level of independence \cite{wang2022survey}. In the Metaverse, the value of users' production and work activities will be confirmed in the form of the unified currency of the platform \cite{belk2022money}. This currency can be used to consume content on the Metaverse platform and can also be exchanged for legal tender in real life at a certain ratio.

\subsection{Key Technologies of the Metaverse}

Metaverse is the combination of the real world, the virtual world, and human society. To build it, many different software and hardware technologies must be used in a wide range of ways. There are six key technologies, including blockchain, human-computer interactivity, electronic games, artificial intelligence (AI), network and computing technology, and the Internet of Things (IoT). Key technologies evolved with the development of the Metaverse. Previous studies \cite{mozumder2022roadmap,ning2021meteverse} believe that there are four categories of key technologies in the Metaverse: communication computing infrastructure, fundamental common technology, virtual reality object connection, and virtual reality space convergence.

\textbf{Communication computing infrastructure.} 1) \textit{Communication technology}: Real-time transmission and real-time interaction are essential because the Metaverse is immersive and low-latency \cite{tang2022roadmap}. 5G has a huge bandwidth, low latency, and excellent dependability \cite{alliance20155g}, making it feasible for real-time communication within the Metaverse. As the newest kind of connectivity, 6G is capable of incredibly quick transmission. Users in the Metaverse can get better input faster thanks to 5G and 6G, resulting in a more immersive experience \cite{lee2022metaverse}. 2) \textit{Internet of things (IoT)}: In the Metaverse, the IoT acts as an information bridge between the real and virtual worlds \cite{fang2021metaverse}. Users can interact with the real and virtual worlds in real-time thanks to IoT \cite{simiscuka2018synchronisation}. Because interactive objects equipped with sensors and smart terminals can quickly collect, transmit, and process data, the Metaverse provides users with an instant and smooth interactive experience. 3) \textit{Other technologies}: Cloud computing has efficient computing capabilities, while edge computing provides data computing services at the edge of the network, providing low-latency services. Quantum communication can ensure the security of Metaverse communication \cite{ning2021meteverse}.

\textbf{Fundamental common technologies.} 1)  \textit{Artificial intelligence (AI)}: It is an intelligence that can imitate human intelligence for learning and creation. AI accesses enough data and calculates rich knowledge through appropriate algorithms. Previous studies \cite{mozumder2022roadmap,ning2021meteverse,yang2022fusing} believe that one of the roles of AI in the Metaverse is to generate digital content, and AI technologies such as computer vision, natural language processing (NLP), and speech recognition are responsible for it. 2) \textit{Other technologies}: Avatar is the digital avatar of users in the Metaverse, which captures our appearance characteristics and reproduces them in the Metaverse. The spatio-temporal-based algorithm determines the time passage of the positioning and movement of the user space in the metaverse, helping the Metaverse run more efficiently \cite{mozumder2022roadmap}.

\textbf{Virtual reality object connection.} 1) \textit{Blockchain}: It is a chain composed of information blocks generated in chronological order \cite{si2019iot, zheng2018blockchain}. Blockchain truly realizes the decentralization of information, and the data is difficult to tamper with. Therefore, blockchain creates a trust mechanism, so that each user can create their own identity, social relationships, and personal property in the Metaverse. "Peer-to-peer" transmission can also make the assets and belongings in the Metaverse flow safely and reliably between users, helping the Metaverse to realize social and economic systems. In the Metaverse, blockchain is applied to transactions, copyright confirmation, etc. 2)  \textit{Other technologies}: Identity modeling is the embodiment of the user's personal data and personality in the Metaverse. Social computing is to analyze social interaction and development from social behavior \cite{mozumder2022roadmap}. Decentralized technologies including distributed computing, storage, and distributed databases, serve the social relations in the Metaverse. Among them, blockchain makes distributed computing and storage safe and reliable \cite{mozumder2022roadmap}.

\textbf{Virtual reality space convergence.} 1) \textit{Human-computer interaction (HCI)}: HCI is the entrance to the Metaverse, which includes virtual reality technology (VR), augmented reality technology (AR), and mixed reality technology (MR). When the users wear VR headsets, VR replaces the sound and picture of the real world with those of the digital world, and thus the rich digital audio-visual experience provides users with an immersive experience. Through device identification and software calculations, AR makes digital images and sounds superimposed on those of the real world, and then users receive the final visual and auditory sensation. MR is a combination of AR and VR \cite{ohta2015poster}. Digital objects and real objects coexist, and interactions between them also exist. 2) \textit{Video games}: Currently, video games are the most intuitive embodiment of the Metaverse, and they serve as a prototype of the Metaverse \cite{ning2021meteverse}. The real-time rendering of video games applies to the user avatar to make it realistic, and the game engine helps with image development, making immersion \cite{mozumder2022roadmap}. 3) \textit{Other technologies}: include brain-computer interaction (BCI), which collects brain signals, analyzes features, and then outputs feedback, so the output is determined by the users' intentions \cite{mozumder2022roadmap}. The feedback can even be a stimulation signal to the brain, allowing users to generate various sensory experiences \cite{lee2021all}. Holographic image generates three-dimensional imaging visible to the naked eye, free of headsets \cite{ning2021meteverse}. The process of BCI is as shown in Figure \ref{fig:BCI}.

\section{Human-Centric Metaverse}  \label{sec:HCM}
\subsection{The Proposal of HCM}

In order to mitigate the existing or potential negative impacts of the practical application of AI, Stanford University, the University of California, Berkeley, and other universities have respectively established human-centric AI (HAI) research institutions. This represents the official introduction of the concept of HAI. The HAI concept emphasizes that AI research and development should benefit human beings and be ethical \cite{lepri2021ethical, bryson2019society}. Yu \cite{yv2021social} proposed that "the Metaverse is a human-centric future amphibious social ecology". He believes that the Metaverse combines virtuality and reality, and thus it could greatly broaden the scope of human exploration. By using "digital clones", human beings can explore in virtual space without being limited by time and space, create the experience, and realize value. Therefore, we can believe that the Metaverse is human-centric. This is the original proposal for the concept of "human-centric Metaverse" (HCM).

\subsection{Why Metaverse Should be Human-centric}

At present, there is a global trend toward the Metaverse, and the concept of a human-centric Metaverse has been proposed \cite{yv2021social}. The question we should consider carefully is: Why should the Metaverse be human-centric? The specific reasons are as follows.

One of the elements of the Metaverse is immersion, which shows that the Metaverse is supposed to serve people's feelings \cite{kozinets2023immersive}. If one day the Metaverse is put into use, in order to make users willing to spend time and effort in the Metaverse, it is necessary to let users have a sense of immersion, so that they can obtain entertainment, make friends, learn something, and gain real-world experience just like in the real world.

The Internet entrepreneur at Tencent said that all technologies must ultimately serve people. Almost all the development of technologies stems from humans' yearning for a better life, so the realization of technologies should meet human needs. The executive director of the Metaverse Industry Committee said, "The idea of Metaverse originated from people's pursuit of a happy life, so it was created to serve people and make life more convenient." Yu \textit{et al.} \cite{yv2021character} believe that the Metaverse is people's imagination of the ultimate medium, and humans are the measure of the medium, so humans should be at the core of the Metaverse. Because the user's "digital avatar" in the Metaverse has a comprehensive connection with the user's senses, the users have the subjective initiative in the Metaverse. This establishes the subjectivity of human beings in the Metaverse.

The key to the rise and sustainability of the Metaverse is whether it can create a new way of production, life, and cognition for human beings. In other words, human feelings limit the scope of application of the Metaverse. Therefore, the development of the Metaverse should meet human needs and bring a better life to human beings \cite{bale2022comprehensive}. The Metaverse should be human-centric, whether this is due to the features of immersion in the Metaverse, the nature of it as a medium, or the necessities of the Metaverse's evolution.

\subsection{What is a Human-Centric Metaverse?}

\begin{figure}[h]
    \centering
    \includegraphics[scale=0.3]{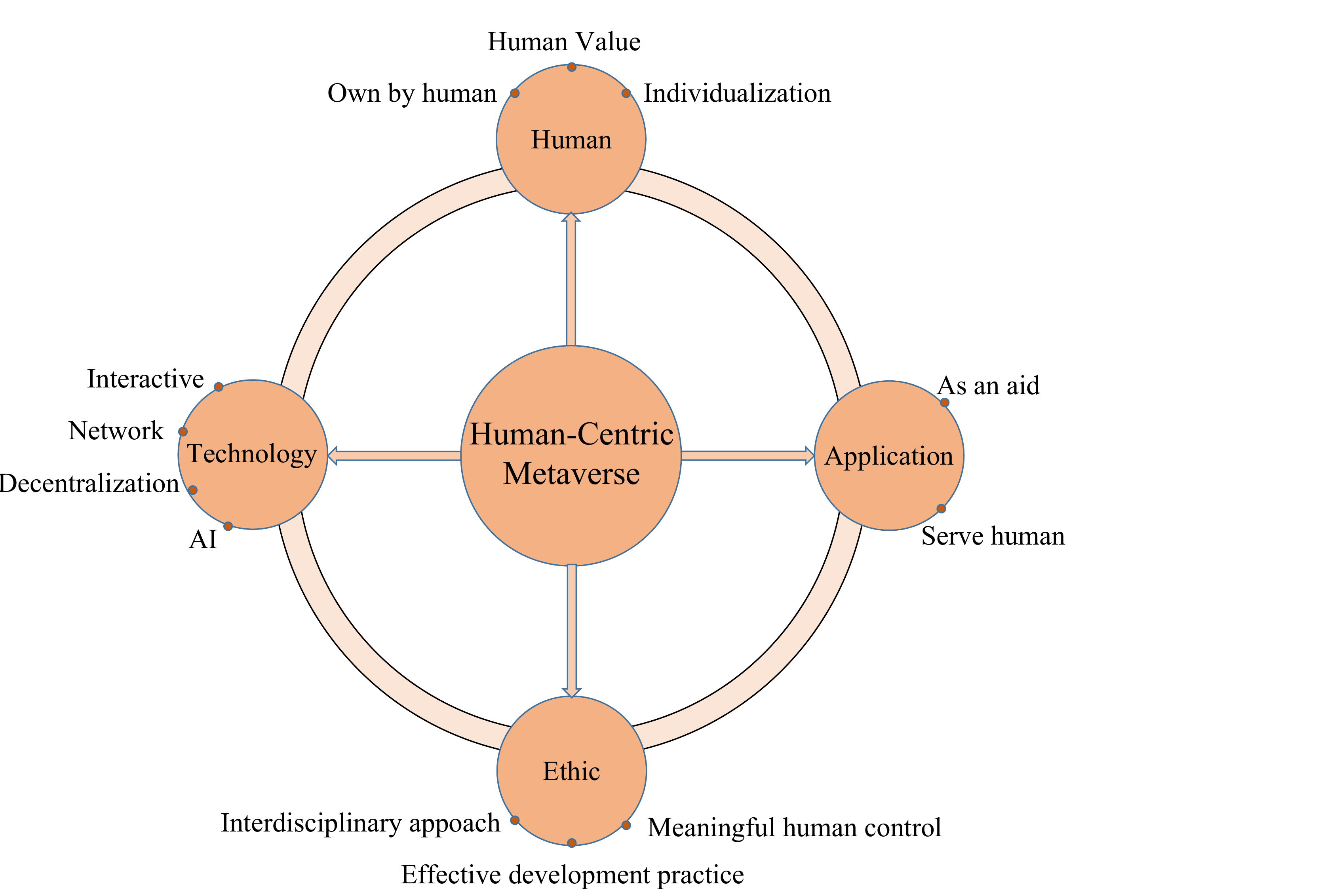}
    \caption{Human-centric Metaverse}
    \label{fig:HCM}
\end{figure}

As shown in Figure \ref{fig:HCM}, the human-centric Metaverse (HCM) should, in our opinion, include human, technological, application, and ethical aspects. Details are provided below.

\textit{\textbf{1) Human aspects:}} HCM refers to the Metaverse owned by users or humans, not by some companies. This Metaverse advocates the will and interests of people, not just the interests of people or companies that promote the design of systems. In other words, the human-centric Metaverse is a Metaverse that takes the realization of human value as the precondition for the realization and sustainable development of the Metaverse and regards the improvement of human progress and well-being as the first principle.

\textit{\textbf{2) Technology aspects:}} HCM is an organic combination of different human-centered technologies. To truly be the ultimate medium, the Metaverse should become an infinitely explorable reality. Therefore, Metaverse needs various technologies to achieve a rich sensory experience, a low-latency network, and a real social life. It also needs to reduce costs to become a service that everyone can accept.

\textit{\textbf{3) Application aspects:}} HCM is an application mode that focuses on reality while supplementing it with virtual reality. It is not a virtual world separate from reality. HCM is combined with various real-world application scenarios, with humans as the main body, and views the ability to build a virtual world as a tool to help people realize various unimaginable experiences in the real world.

\textit{\textbf{4) Ethical aspects:}} HCM is to develop an ethical and responsible Metaverse by combining interdisciplinary methods, effective development practices, and other methods and utilizing meaningful human control to ensure the rights and interests of fairness, human privacy, ethics, human decision-making power, and other aspects in the development of human-centric Metaverse.

\textbf{The core of HCM is human.} Through the mutual support of humans, technology, application, and ethics, HCM can build a world that focuses on human services and always supports the user experience \cite{zeb2022industry}. This Metaverse gives people the right to create, not just consume. It creates a more accessible, more flexible, more experiential, and more effective surreal scene for human beings through the mixed reality of virtual and real integration. In this Metaverse, everyone is the creator and god of their world. The Metaverse can change people's minds and carry out various personalized operations. That is, all services of the Metaverse are carried out around the subject of human beings.

\textbf{The characteristics of HCM.} In our opinion, there are several characteristics of HCM. i) Highly accessible and inclusive. The HCM is the second home created for all and is highly inclusive, which means that the Metaverse can accept the landing and use of all kinds of people. Whether the person is healthy or not, the service of the Metaverse is accessible to them. In addition, everyone, whether healthy or disabled, will be affected by digital accessibility. The HCM has a richer and better digital access mechanism. For example, a single controller designed for the disabled without one hand will also play a significant role when one of the two handheld controllers is affected. ii) Be able to understand user behavior fully. In the Metaverse, everyone will have their own personal role, and people will transmit and show different virtual scenes of the Metaverse. In order to ensure user experience, the HCM needs to understand this new user behavior and expectation deeply. The HCM can better maintain the human-centric user experience by learning, predicting, and integrating human behavior and reasonably allocating the technology and resources of the Metaverse \cite{balica2022metaverse}.

\section{Key Technologies of HCM} \label{sec:Technologies}

The Metaverse is realistic and immersive, giving people the same realistic experience as the real world. But beyond the real world, people are free to explore and create in the Metaverse \cite{yv2021social}. As shown in Figure \ref{fig:technologies}, the Metaverse can use various technologies to achieve full-sensory simulation, and real-time interaction, as well as diverse choices.

\subsection{Sensory Feedback}

VR, AR, brain-computer interaction (BCI), and holographic projection are responsible for providing sensory feedback to users.

\textbf{\emph{1)  Audio-visual feedback:}} VR and AR devices can cover the user's hearing and vision, bringing audio-visual immersion. Holographic projection can generate virtual three-dimensional images visible to the naked eye. Users can interact with the Metaverse more conveniently without wearing heavy equipment. 

\textbf{\emph{2)  Other sensory feedback:}} BCI can also simulate nerve impulses by sending electrical signals to the brain, giving people a sensory experience. Meanwhile, BCI can successfully monitor the user's thoughts and actively assist the user in completing what they want to do and how to do \cite{lee2021all}.

\textbf{\emph{3)  What can be improved:}} It is worth noting that the current VR and AR technologies have only achieved audiovisual simulation. To truly take the concept of a human-centric Metaverse a step further, BCI should also be implemented to allow users to have a sense of touch and taste in the Metaverse, so as to achieve the user's full sensory simulation \cite{yv2021social,lee2021all}.

\subsection{Faster and Better Experience}

The Metaverse is an immersive environment, and users should have real-time interactions with digital items and dynamic communication \cite{jaynes2003metaverse}. So whether the Metaverse is "fast" enough determines whether users can have a good immersion.

\textbf{\emph{1)  Communication technology:}} At present, the minimum standard of communication technology required by Metaverse is 5G. With low latency, network services will not suffer from slow content loading, video freezing, and other unsmooth phenomena like 4G networks. 6G can achieve extremely fast transmission and is expected to help realize the real interconnection of everything and provide a network foundation for the Metaverse \cite{tang2022roadmap}.

\textbf{\emph{2)  Computing:}} Cloud computing provides powerful computing capabilities, while edge computing provides low-latency computing services, and the combination of the two in the Metaverse can obtain low-latency and powerful computing capabilities. This enables faster data transmission in the Metaverse, offering users a real-time immersive experience \cite{cai2022compute}.

\textbf{\emph{3) The Internet of Things:}} Thanks to 6G and the internet of things (IoT), in the future Metaverse, human life will benefit from intelligent interaction anytime, anywhere \cite{li2022internet}. Such interactions allow users to more easily use tools to solve problems in the Metaverse. These prove that the Metaverse is human-centric.

\textbf{\emph{4)  What can be improved:}} However, in order to enable IoT technology to truly bring users smooth and real interactions, efforts are still needed in data collection and transmission. It relies on more convenient interactive devices with better sensors, such as nanosensors \cite{ning2021meteverse}. Besides, 5G is not fast enough for the Metaverse \cite{lee2021all}, while 6G transmission capacity is greatly improved, and it may meet the standard of the Metaverse.
 
\subsection{Users' Social Interaction}

Technologies such as avatars, identity modeling, blockchain, and social computing serve users’ social interaction. 

\textbf{\emph{1)  User's avatar:}} Avatars are digital avatars of users in the Metaverse. 3D scanning technology is now available to make avatars that reflect the appearance of people as realistically as possible. Rigorous identity modeling allows user data to more accurately reflect the characteristics of users, providing a basis for users' social interaction in the Metaverse. Besides, the GAN can automatically generate a 2D virtual avatar \cite{lee2021all}, allowing users to choose a virtual avatar according to their preferences. 

\textbf{\emph{2)  Real-time interaction:}} The real-time rendering technology of video games can also be applied in the Metaverse. Real-time rendering \cite{mozumder2022roadmap} of the user's avatar to make it more realistically reflect the user's demeanor and movements, making the communication between users more real and full of immersion.

\textbf{\emph{3)  Social computing:}} Social computing \cite{ning2021meteverse} studies the social development trend of the Metaverse by collecting users’ data so that the Metaverse can better serve people. The blockchain \cite{yang2022fusing} creates a trust mechanism, which enables the huge amount of user data in the Metaverse to be properly preserved. Blockchain also ensures the safe flow of messages and digital items (such as digital currency) in the Metaverse, providing security for users' social relationships and economic systems. If quantum communication can be implemented, the communication between users will become more secure, and the concept of human-centric will go further.

\textbf{\emph{4)  Economic:}} Metaverse will deploy transaction methods for users. It will be built on the basis of the blockchain, and users can use cryptocurrencies that are connected with real-world currencies to produce or consume in the Metaverse \cite{yang2022fusing}. Cryptocurrencies will become the mainstream payment method in the Metaverse \cite{laeeq2022metaverse}. And the decentralization of blockchain makes the virtual assets and the identities of users reliable \cite{yang2022fusing}. Blockchain and cryptocurrencies can guarantee the development of commerce in the Metaverse, allowing users to create economic value.

\textbf{\emph{5)  What can be improved:}} In the Metaverse, lots of users are trading every second, which means blockchain nodes need to process a large number of transactions in a short time. As a result, related improvement should be introduced \cite{yang2022fusing}. Cryptocurrencies are poorly regulated, and they can be used anonymously or even avoid censorship, which allows criminals to take advantage of them and also makes cryptocurrencies a new means of fraud \cite{mackenzie2022criminology}.

\subsection{People's Acceptance}

Because Metaverse is the landing site for the new generation of the Internet \cite{mozumder2022roadmap}, its cost and services must be acceptable to the general public, as most people spend a significant amount of time on the Internet every day.

\textbf{\emph{1)  Lower cost:}} The game engine is the core technology of video games, it makes the development of the game unnecessary to start from scratch, and thus reduces a lot of repetitive work. The application of the game engine in the Metaverse can make it easier to create beautiful and realistic pictures, and reduce the cost of image development \cite{ning2021meteverse}. What's more, after accepting massive user data and machine learning, AI can independently generate virtual world content, and this can greatly reduce the cost of users in the Metaverse. It makes the Metaverse cheaper to implement and increases its acceptance among users. 

\textbf{\emph{2)  Better service:}} As one of the key technologies of Metaverse, AI technology can be used to create digital humans and generate content. Digital humans can serve as guides in the Metaverse, and by using NPL, digital humans can understand complicated conversations regardless of users' accents or backgrounds \cite{huynh2023artificial}, thus promoting the Metaverse in the world. 

\textbf{\emph{3)  More options and easier life:}} As the Metaverse grows, people spend more time in the Metaverse, so users beautify their avatars as much as possible, making them look nice and attractive \cite{laeeq2022metaverse}. Avatar is the digital avatar of users, so digital decorations and clothing are easier to realize than those in the real world. Eliminating the time cost of makeup and the money cost of buying beautiful clothes in reality, avatars in the Metaverse will be very attractive to users. In the future, users can even create hyperreal items easily with the help of AI, and this undoubtedly can provide users with more choices \cite{huynh2023artificial}. 

\textbf{\emph{4)  The mixed reality:}} The mixed reality provides easier lives and equal opportunities for users. In the Metaverse, user avatars will be able to teleport \cite{mystakidis2022metaverse} without restriction. Imagine that one day in the office, a good idea occurred in your mind, and you could immediately have a meal with entrepreneurs from other countries, within a few seconds. This saves a lot of money and time on transactions, making life easier. At the same time, the mixed reality of the Metaverse makes many remote services possible. At that time, it will be possible for users to have equal opportunities for work and education regardless of location. For example, social mixed reality in the Metaverse can contribute to the democratization of education \cite{mystakidis2022metaverse}.

\textbf{\emph{5)  What can be improved:}} It should be noted that AI is users' tool to create, and thus ethic issues should be examined by Metaverse managers and users to reduce risks \cite{huynh2023artificial}. In general, AI should be cheap to meet users' acceptance.

\section{Key Applications of HCM}  \label{sec:Application}

Under the influence of the global epidemic, people's work modes and living and entertainment styles have changed significantly due to the COVID-19 control policy, from offline physical offices and entertainment to online virtual \cite{chen2021influence}. Through virtual and real integration, the Metaverse has the potential to profoundly alter the organization and operation of the existing society. It is not a substitute for real life. However, it will create a new way of life with virtual and real two dimensions, hastening the birth of a new social relationship of online and offline integration and providing new vitality to the actual economy from the virtual dimension. As a result, the Metaverse is in an early stage of development and has a wide range of applications.

As shown in Figure \ref{fig:application}, the Metaverse has a wide range of applications. These application scenarios also contain the concept of human-centric. Details are present in the following subsections.

\subsection{Education Metaverse}

In the post-epidemic era, the Metaverse gradually showed its unique role in education. In natural science and mathematics, Metaverse can provide technical support. Metaverse can be used for language learning and communication in the arts and humanities. In the field of archaeology, the use of Metaverse can also provide students with independent virtual mining opportunities \cite{tlili2022metaverse, boyes2018industrial}. In this kind of educational application, the concept of being human-centric or student-centered is embodied. The Metaverse uses augmented reality technology to make the virtual classroom more vivid, enabling students to experience unique experiences. For example, the Cruscope's virtual-tee as shown in Figure \ref{fig:HCM_Reality}(a), can enable students to examine the body interior, just like in the anatomy room. By using virtual world technology to create virtual classes, students can take classes online without going out. For example, the video conferences systems such as Zoom, Teams, Tencent Conference, and the classroom map in various 3D maps of Zepeto\footnote{https://web.zepeto.me} play the role of the classroom in the post-epidemic era and realize remote non-face-to-face teaching \cite{kye2021educational}. The human-centric Metaverse serves students by using augmented reality, artificial intelligence, and other technologies to make learning more vivid, convenient, and comprehensive.

The Metaverse gives learners more opportunities to experience, explore, learn, teach, cooperate, and interact with others in the new world \cite{hwang2022definition}. In the Metaverse, learners can learn in a virtual and dangerous environment, experience the real world that they usually have no chance to access, learn various knowledge and skills that cannot be learned in the real world due to objective reasons, and learn content that is irrelevant to their profession or specialty. The foundation for realizing this is the Metaverse, which is supported by the technology of the Metaverse and centered on the concept of human-centric and student-centered.

\subsection{Medical Metaverse}

In medicine, the Metaverse also builds scenes and provides various services around people. Among them, virtual reality technology provides adequate support for disaster medicine \cite{wu2022scoping}. The construction of post-disaster scenarios through virtual reality technology will help medical workers cope with this chaotic environment calmly and efficiently and will not shrink their skills because of the rarity of such events so that medical workers can face various disaster scenarios easily. In clinical medicine, as shown in Figure \ref{fig:HCM_Reality}(b), the intense immersion of the Metaverse can allow experts to conduct real-time remote guidance through the online construction of surgical scenes \cite{thomason2021metahealth}. Through the Metaverse, time and space constraints have been broken, enabling experts from all over the world to brainstorm and face problems and challenges together. The use of the Internet of Things technology and the embedded devices that integrate information and objects give the Metaverse excellent development potential in medical care \cite{yang2022expert,skalidis2022cardioverse}. For example, the UK has developed wireless sensor networks. Sensors embedded in cabinets, shoes, and other places can help the elderly when necessary. Applying Metaverse technology in surgery simulation, patient care, health management, and other aspects can effectively improve patients' experiences. In addition, the HCM also plays a massive role in the education of medical students \cite{koo2021training}. Using AR and VR technology \cite{almarzouqi2022prediction}, medical students can enter the human body. By building natural scenes, medical students can have a comprehensive perspective and immersive operation experience \cite{thomason2021metahealth}. In the above applications, doctors, nurses, patients, and medical students are all roles that people play in the medical field. The Metaverse is the embodiment of the human-centric concept by providing different services around different roles in different fields.

\subsection{Tourism Metaverse}

The arrival of the epidemic has powerfully hit the tourism industry. Moreover, the traditional tourism industry also has some common problems. For example, the travel route of tourists is determined passively, and there is no incentive to encourage tourists to continue to browse, and it is challenging to promote other industries \cite{wei2022gemiverse}. However, the popularity of the Metaverse and the human-centric concept have brought some changes to the tourism industry, which has spawned another way of tourism, that is, smart tourism \cite{dwivedi2022metaverse}. Smart tourism means regarding tourists as the center, and then we can enjoy the fun of tourism without leaving our homes, by using artificial intelligence, big data, and other related technologies of the Metaverse\footnote{https://en.wikipedia.org/wiki/Smart\_tourism}. For example, in the travel of scenic spots and historic sites, enhancing cultural heritage sites through immersive MR and AR experiences can add value to the whole user's participation, especially in combination with customization based on each customer's preferences, experience, and interaction with historical projects in the real world \cite{buhalis2022mixed}. 

Figure \ref{fig:HCM_Reality} (c) shows this specific application The Natural History Museum has changed the form of the fossil exhibition through the applications of AR. Through the application, visitors can cover the mobile phone on the fossil and see the animals "resurrected in situ", and the background related to the fossil will also be vividly displayed. Similar technologies will be applied to exhibitions in scenic spots. Try to imagine that the ancient buildings or historical figures in front of us become vivid and interact with us enthusiastically. We can even travel in space by using Metaverse's technology. In the real world, space travel is an expensive business activity. However, in the Metaverse, we can achieve immersive space travel through virtual technology \cite{zaman2022meet} and customize our interstellar travel. 

The human-centric Metaverse provides a solution for the tourism industry, featuring real-time, spatial, immersive, customized, 3D rendering, and dynamic interaction. However, the combination of the two is not limited to the use of 3D rendering equipment such as XR to promote scenic spot projects but also needs to use the multidimensional digital virtual environment and open cultural and creative ecology of the universe to build an inclusive, free development, tourist-centered tourism virtual world, and achieve all-round service and industrial innovation.

\subsection{Industrial Metaverse}

As shown in Figure \ref{fig:HCM_Reality}(d), Huang \textit{et al.} cheated people worldwide with a 14-second digital double\footnote{https://blogs.nvidia.com/blog/2021/08/11/omniverse-making-of-gtc/}, at the NVIDIA press conference. This 14 seconds clip is a successful marketing of the Industrial Metaverse. The industrial Metaverse was born from the industrial internet. The industrial internet is a combination of physical equipment and information technology \cite{sisinni2018industrial}. However, the industrial Metaverse is the integration of the Metaverse and industry, which closely integrates the real physical and virtual digital worlds\footnote{https://www2.deloitte.com/cn/en/pages/technology-media-and-telecommunications/articles/metaverse-report.html}. Metaverse uses its unique hyper spatio-temporal characteristics to realize virtual verification design, plan and optimize the manufacturing process of the product life cycle, and solve the problems of the long product test cycle and unstable manufacturing process. In addition, the industrial Metaverse pays more attention to the collaboration between virtual and real space. It emphasizes mapping and broadening the operations that can be realized by the physical industry in the virtual space, guiding the efficient operation of the physical industry through the collaborative work and simulation operations in the virtual space, empowering all links and scenes of the industry, enabling industrial enterprises to achieve the goal of reducing costs and improving production efficiency, promoting efficient collaboration within and between enterprises, helping the high-quality development of industry, and realizing the further upgrading of intelligent manufacturing. In the era of the industrial Metaverse, most physical products purchased by consumers come with digital twins connected to the Internet of Things in real time and forever. All kinds of physical information and consumer opinions in the process of product use can be fed back in real-time to brand owners, research and development institutions, manufacturers, and other stakeholders in real-time through this digital twin, and those can be combined to improve and perfect the product \cite{jiang2021industrial}.

The industrial Metaverse is centered on three main entities: products, businesses, and consumers. It realizes the human-centric concept and propels the industry to new heights by leveraging comprehensive digitalization, the support of blockchain production relations, immersive communication, and the power of ubiquitous AI technologies.

\section{Open Problems and Opportunities} \label{sec:Problems}

It is undeniable that this is the most subversive and innovative era. The intergenerational cycle of the new generation of information technology innovation with digitalization, networking, and intelligence as its core has been significantly shortened, and the application potential has been released in a fission way, which is triggering the scientific and technological revolution and industrial transformation at a faster speed, a broader scope, and a deeper degree. However, despite the incredible pace of change, there are still many challenges for realizing the human-centric Metaverse.

\textbf{Cooperation.} In order to realize the human-centric concept, there is a lack of effective interdisciplinary collaboration. For example, the personnel in charge of various technologies in the Metaverse lack the knowledge reserves of other disciplines. A more effective communication language between different technologies is also required. More effective cross-disciplinary collaboration methods are also required. For example, some human-computer interaction professionals continue to use traditional methods for non-AI systems when developing AI systems, despite widespread acceptance of appropriate methods from other disciplines. Furthermore, some people interpret the term "human-centric" differently. For example, many AI developers believe that the problems that cannot be solved by human-computer interaction, are now solved by AI technology, so the human-computer interface does not need to consider user experience, and professionals such as those in human-computer interaction frequently participate in the project after the product requirements of the AI project are defined. Their collaboration frequently fails, resulting in the project's failure. As shown in Figure \ref{fig:cooperation}, we believe that we can solve the problem of cooperation from three aspects: organization, society, and team.

\textbf{Network and computing power.} The immersive experience of the Metaverse can not be separated from the support of computer graphics and image computing power, nor can it be separated from low-latency network services. At the same time, it also needs strong AI computing power and ubiquitous network connections \cite{chang20226g}. The ability of networks and computing power directly determines the depth and breadth of the application of the Metaverse. However, there are still many problems with computing power and the network. i) The problem of network and computing resources. The network technology system and architecture are built around people, and the computing technology system and architecture are built around data, which means that there are differences between the two systems at the beginning of construction. ii) The contradiction between ubiquitous networks and low-latency networks. Low-delay networks require a direct connection at both ends of communication nodes as much as possible, while ubiquitous networks require systematic design. If all flat mesh networks are used to achieve ubiquitous full coverage, the cost of network construction, operation, and maintenance will be huge.

\textbf{Interactive technologies.} There are many users playing in the Metaverse at the same time, so real-time interaction between users and virtual objects and multi-user interactions must be realized, and such interactions should be consistent for all users. The users' experience may be ruined because the processing of dynamic events may cause delays or the information received by users is difficult to synchronize \cite{lee2021all}. VR headsets are currently relatively heavy, and wearing them for an extended period of time can easily cause shoulder and neck strain. Furthermore, VR devices frequently cause nausea and motion sickness, and likely psychologically make users outliers. Through interactive technologies, lawbreakers may cause physical and mental harm to users. Due to the immersion of the interaction technologies of the Metaverse, malicious programs can implant uncomfortable virtual content, such as dizzying pictures or violent pictures, thereby affecting the user's feelings \cite{lebeck2018towards}. Because of the immersion, violent incidents in the virtual world will provide users with traumatic experiences without causing physical harm \cite{mystakidis2022metaverse}. In another case, if a malicious user in the AR world makes a startling gesture towards other users, this may distract the users from what they are doing, causing an accident \cite{lebeck2018towards}.

\textbf{Privacy and security.}  Metaverse provides users with immersive experiences that are based on data collected from the real world. Sensors attached to users can monitor and collect a vast amount of physical and biometric data that might be sensitive to the users, such as their personal aspects and sexual preferences. The Metaverse collects and transmits personal information almost constantly during the unprecedented immersion and interactivity by using sensors to scan and monitor the users' surroundings. In such a case, the risk of privacy violation has increased significantly \cite{wang2022survey,di2021metaverse}. Data collection and data sharing by interactive devices (such as AR) pose a great privacy risk \cite{wang2022survey,mystakidis2022metaverse}. According to the study \cite{wang2022survey}, Metaverse managers may collect the user's biometric information on the terminal device worn by the user, such as eye movement, facial expression, brain wave, and so on, and then meet the user's expectations based on this information, bringing users more extreme immersion. This information may be shared among Metaverse companies to improve Metaverse services. However, the relevant storage, protection, destruction, and other processes have not been perfected, and there is a hidden danger of being stolen. Identity theft is possible in the Metaverse. Hackers can use system flaws to commit identity theft, exposing private data such as assets, passwords, and social lives to hackers. Not only that, after successful identity theft, hackers can even use the user's avatar to steal more privacy from the user's relatives and friends.

\textbf{Digital ethics.} Although digital ethics is not dissimilar to traditional ethics, there are risks associated with large-scale unintentional or intentional unethical behavior. With the uncontrolled technological development of the Metaverse, the previous studies mostly highlighted the co-existing opportunities and threats for humanity. Given the potential scale of impact, it is necessary to start exploring ethical questions in the Metaverse. Forming and disseminating new socio-humanitarian rationality is a necessary condition for the successful development of the Metaverse, which will ensure control over the actions and activities of users. For example, users should not be distinguished based on their ethnicity, language, religion, socioeconomic status, or other factors, so that everyone can express their rights in the Metaverse. Meanwhile, algorithms should not create prejudice based on erroneous connections made by an unethical individual. Thus, identifying discrimination in algorithms is a crucial and required effort. A well-managed Metaverse should comply with ethical standards to extend the reach of its services in the next generation of the Internet.

\section{Conclusion} \label{sec:conclusion}

In this paper, we briefly describe the Metaverse's relevant concepts, characteristics, and technical support. In addition, the concept of the human-centric Metaverse is proposed. Specifically, we introduce the concept of human-centric, what is the human-centric Metaverse, and why the Metaverse should be human-centric. Further, we detail the concrete embodiment of the human-centric concept in the Metaverse technology, the combination, and application of the human-centric Metaverse and reality. Finally, we discuss and summarize the challenges faced by the current human-centric Metaverse construction. We hope this paper can inspire more open research in the field of Metaverse, promote the deep implementation of the human-centric concept in the Metaverse, and provide researchers with new ideas for the development of the Metaverse and the iteration of the new Internet.

\begin{acks}
    This research was supported in part by Guangzhou Basic and Applied Basic Research Foundation (No. 202102020277), Fundamental Research Funds for the Central Universities of Jinan University (No. 21622416), National Natural Science Foundation of China (Nos. 62002136 and 62272196), Natural Science Foundation of Guangdong Province (No. 2022A1515011861), the Young Scholar Program of Pazhou Lab (No. PZL2021KF0023), and Guangdong Key Laboratory for Data Security and Privacy Preserving. Dr. Zhenlian Qi is the corresponding author of this paper. 
\end{acks}

\newpage

\bibliographystyle{ACM-Reference-Format}
\bibliography{paper}

\appendix

\renewcommand\thefigure{\Alph{section}\arabic{figure}}
\renewcommand\thetable{\Alph{section}\arabic{table}}
\section{Appendix} \label{sec:appendix}

\begin{figure}[H]
    \centering
    \includegraphics[width=0.47\textwidth]{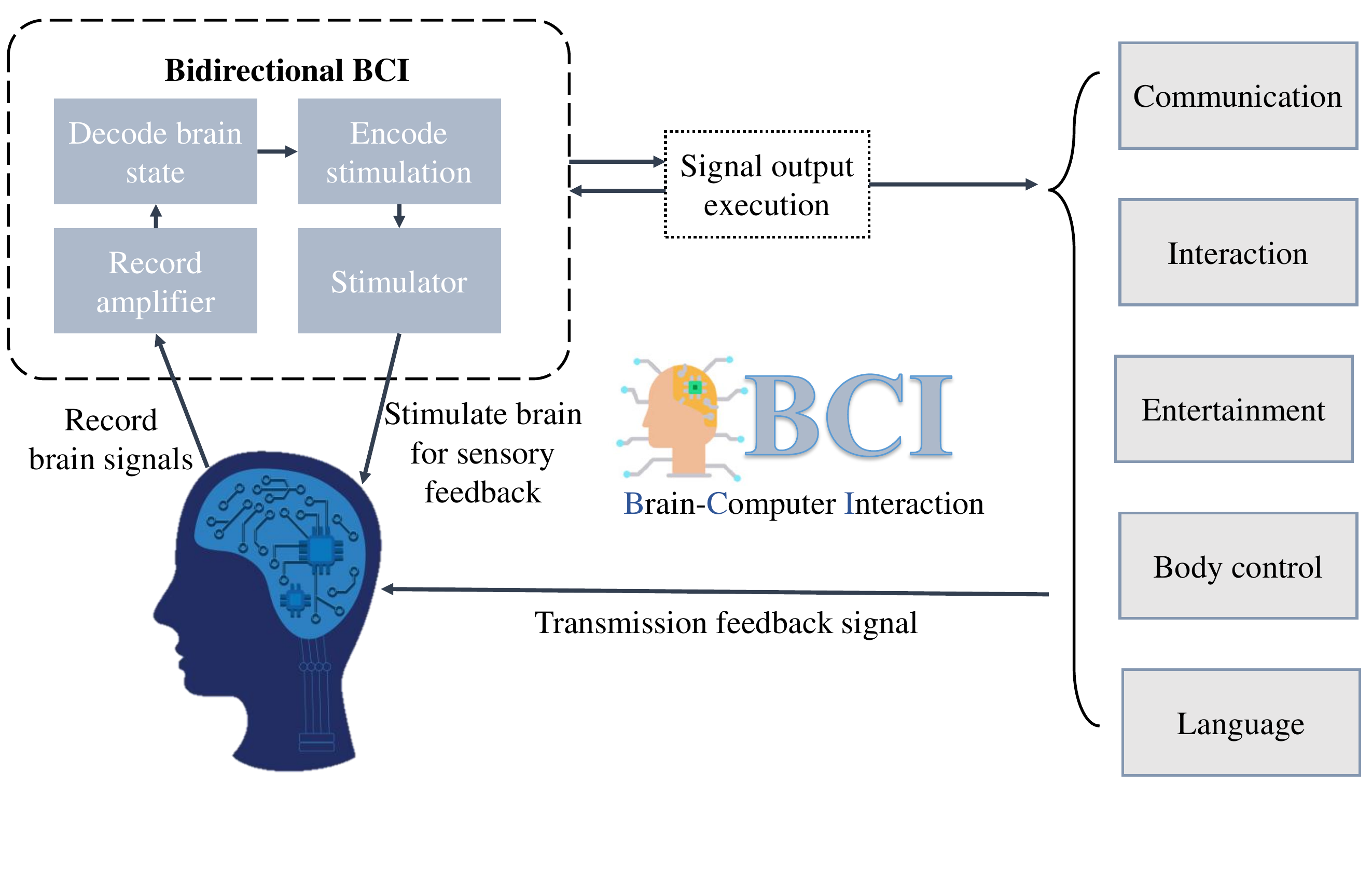}
    \caption{The process of brain-computer interaction.}
    \label{fig:BCI}
\end{figure}

\begin{figure}[H]
    \centering
    \includegraphics[scale=0.32]{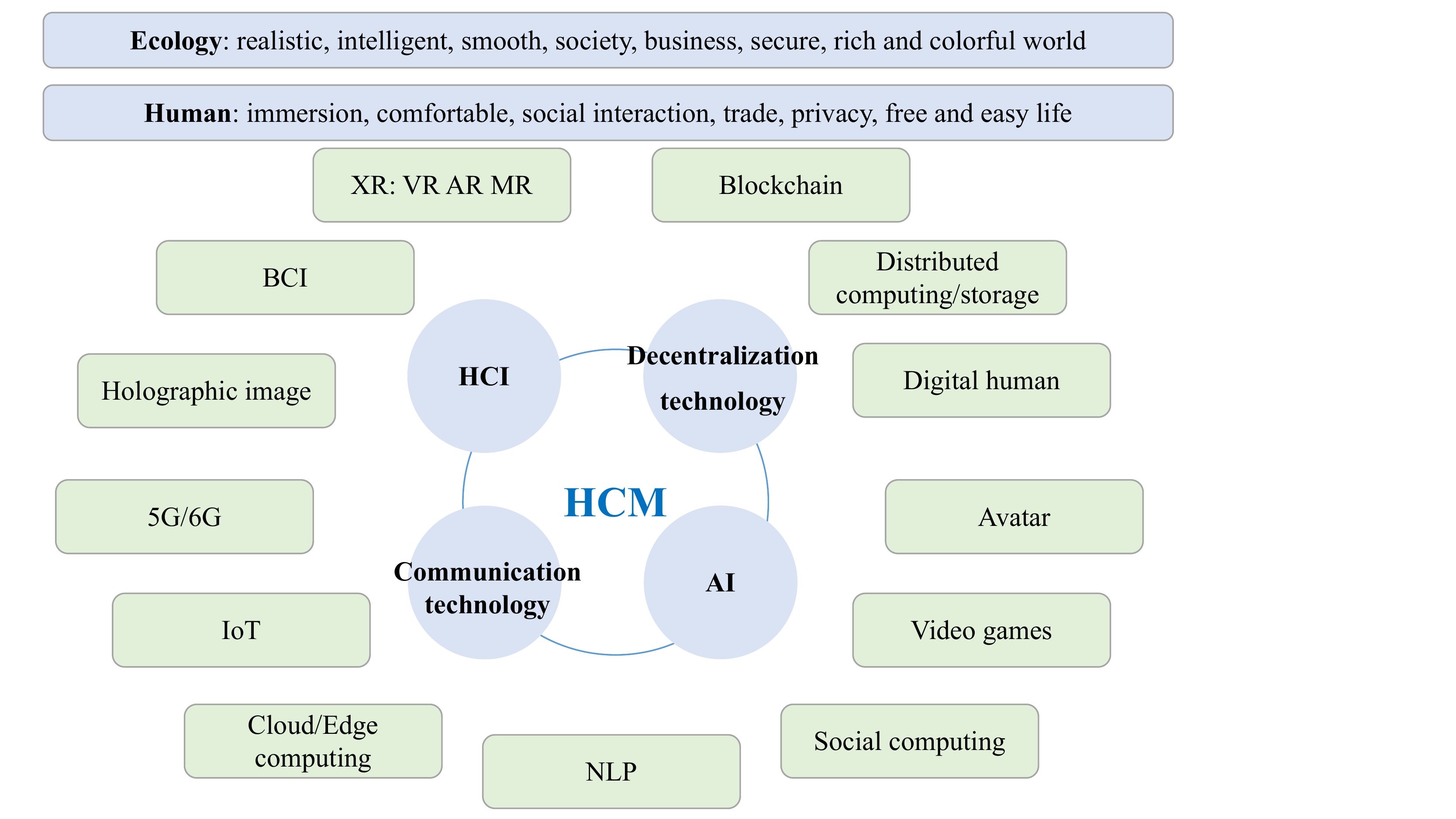}
    \caption{Key technologies of human-centric Metaverse.}
    \label{fig:technologies}
\end{figure}

\begin{figure}[H]
    \centering
    \includegraphics[scale=0.3]{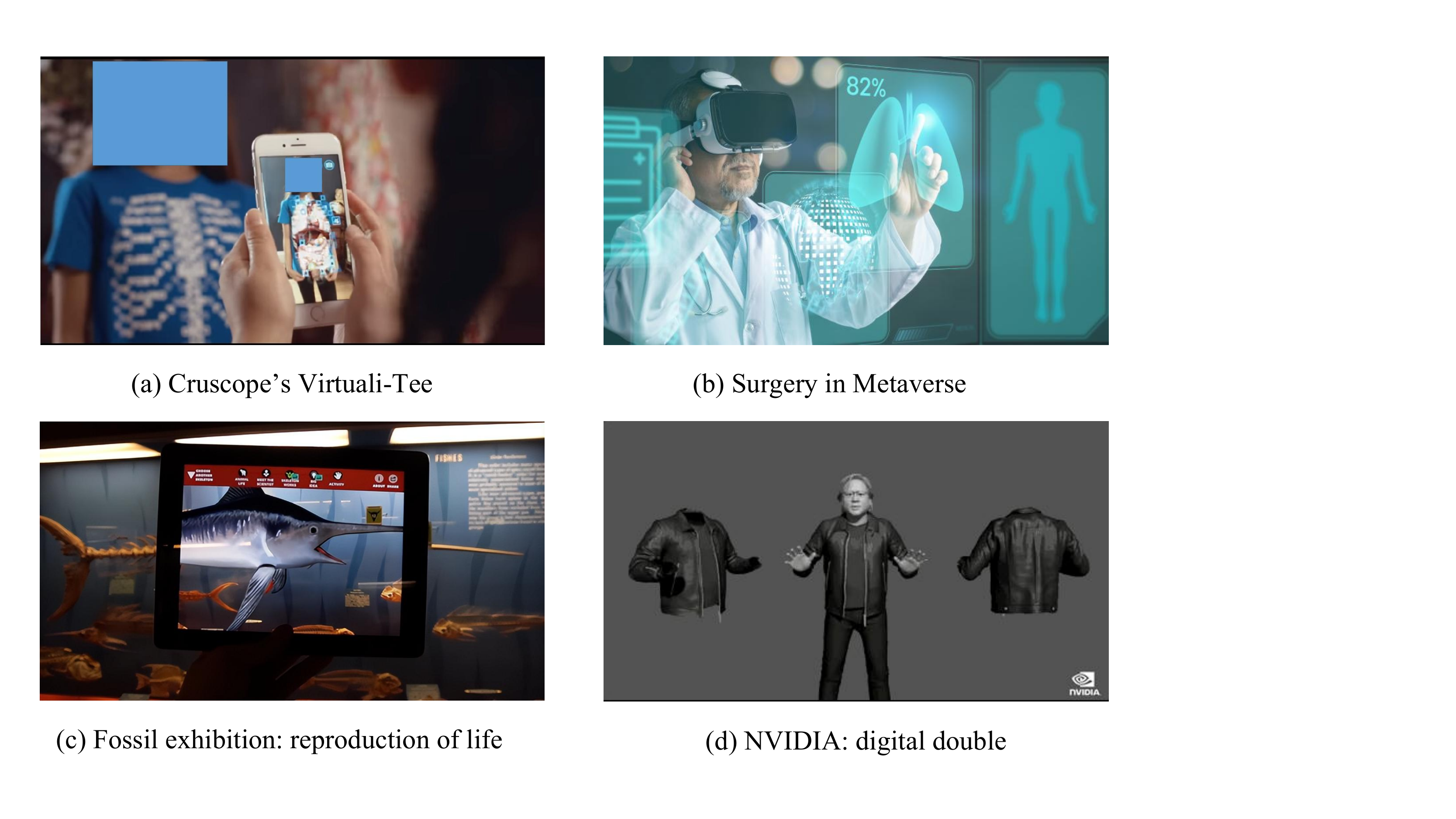}
    \caption{Human-centric Metaverse in education \protect\cite{kye2021educational}, Medical\protect\footnote{https://www.cognihab.com/blog/how-practical-is-surgery-in-metaverse/}, Tourism\protect\footnote{https://www.yuanyuzhouneican.com/article-120350.html} and Industry\protect\footnote{https://news.creaders.net/us/2021/08/12/2386114.html}.}
    \label{fig:HCM_Reality}
\end{figure}

\begin{figure}[H]
    \centering
    \includegraphics[width=0.35\textwidth]{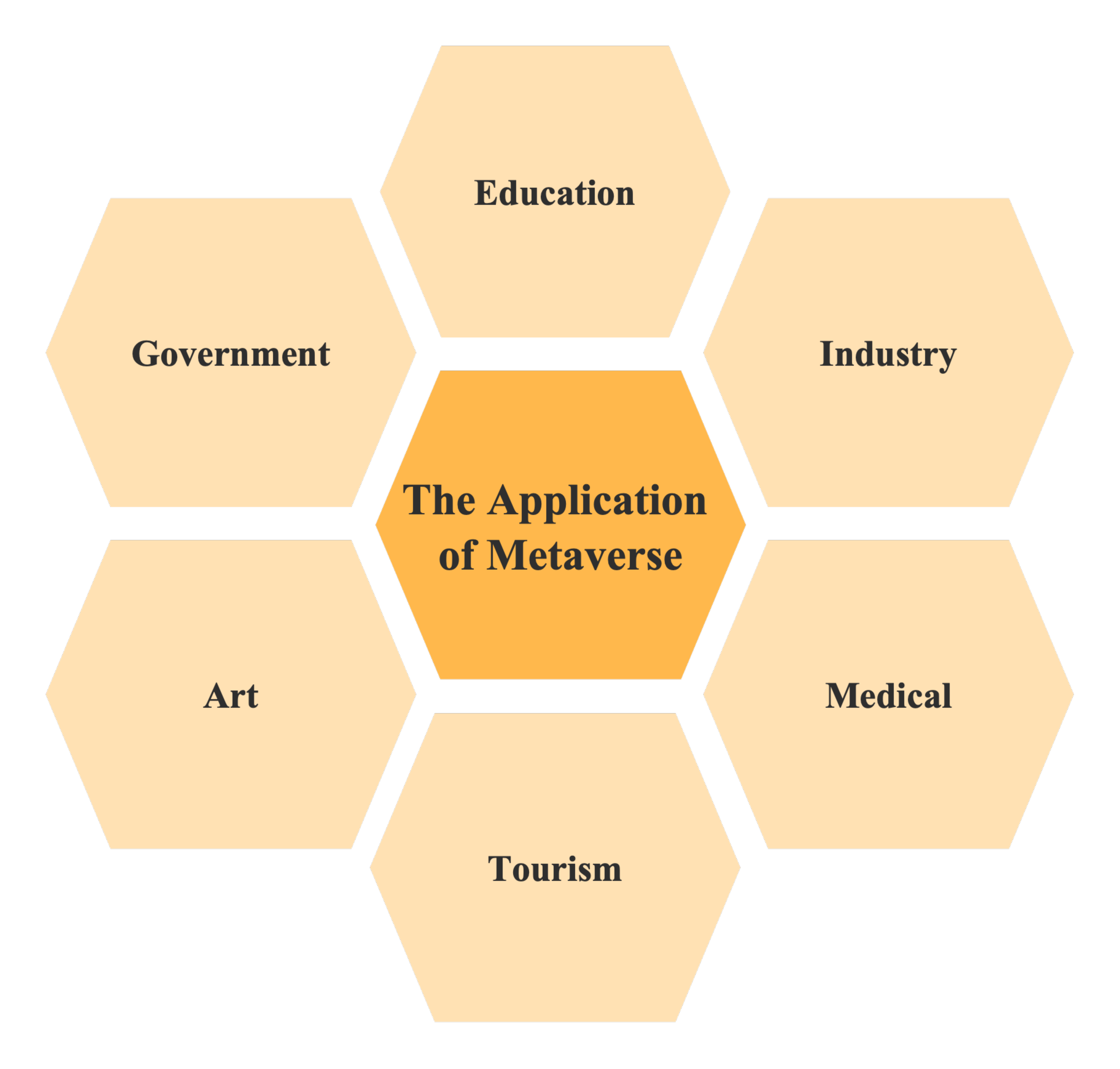}
    \caption{The applications of Metaverse.}
    \label{fig:application}
\end{figure}

\begin{figure}[H]
    \centering
    \includegraphics[scale=0.3]{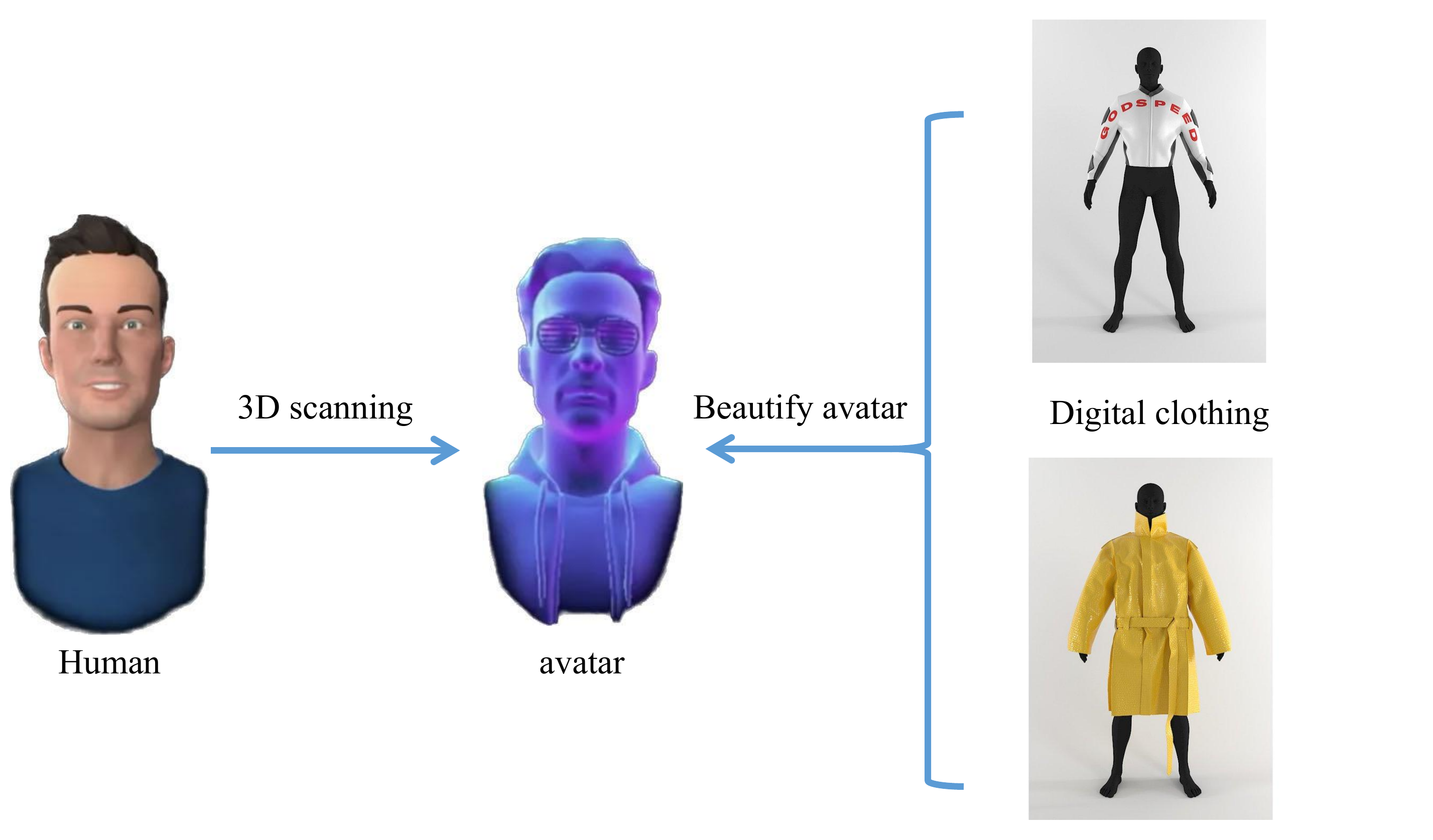}
    \caption{User's avatar\protect\footnote{https://www.trendhunter.com/trends/expressive-avatar, https://visualatelier8.com/carlings-launches-a-digital-only-collection/}}
    \label{fig:avatar}
\end{figure}

\begin{figure}[H]
    \centering
    \includegraphics[scale=0.28]{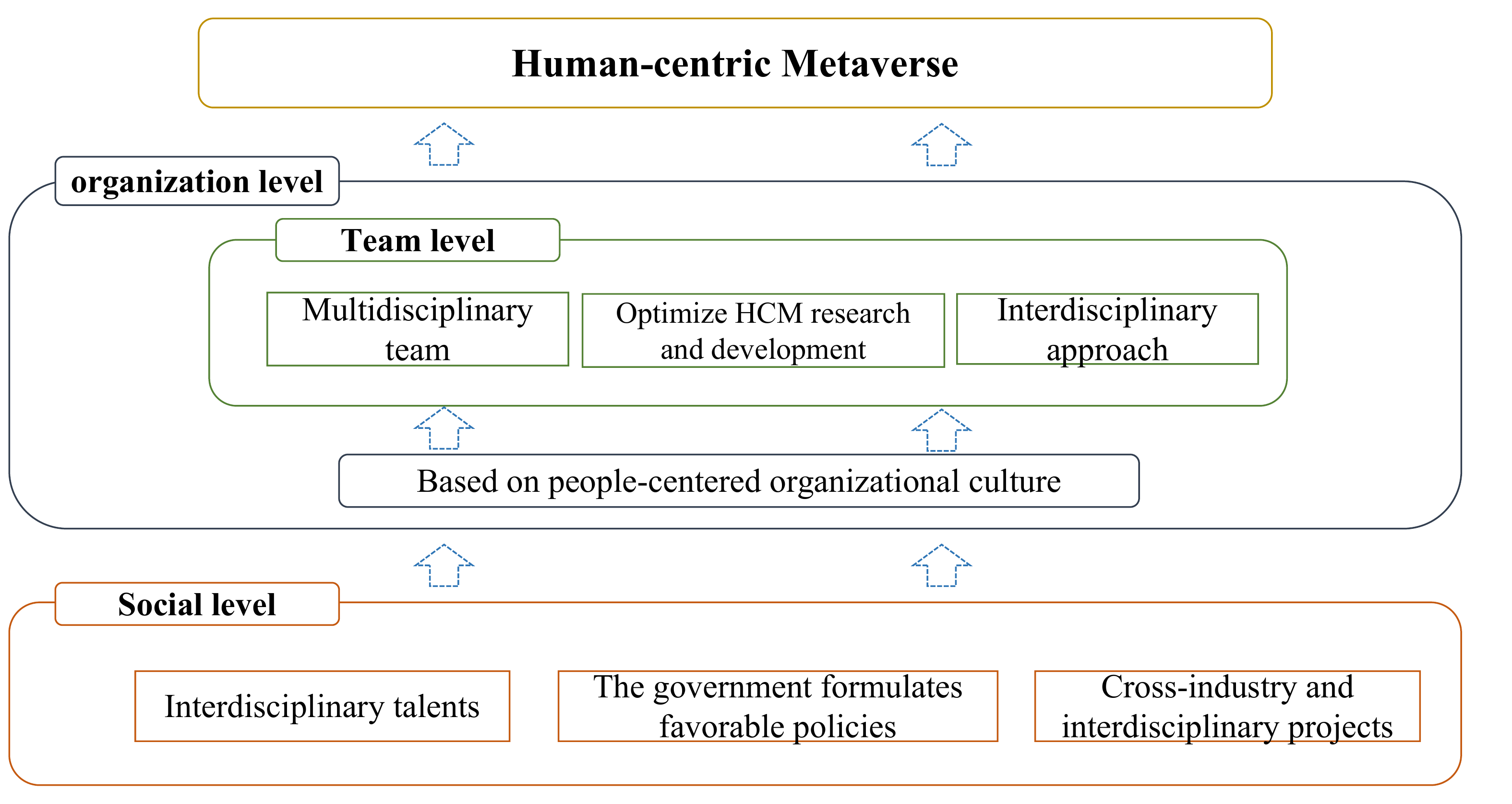}
    \caption{Strategies for solving cooperation.}
    \label{fig:cooperation}
\end{figure}

\begin{figure}[H]
    \centering
    \includegraphics[scale=0.36]{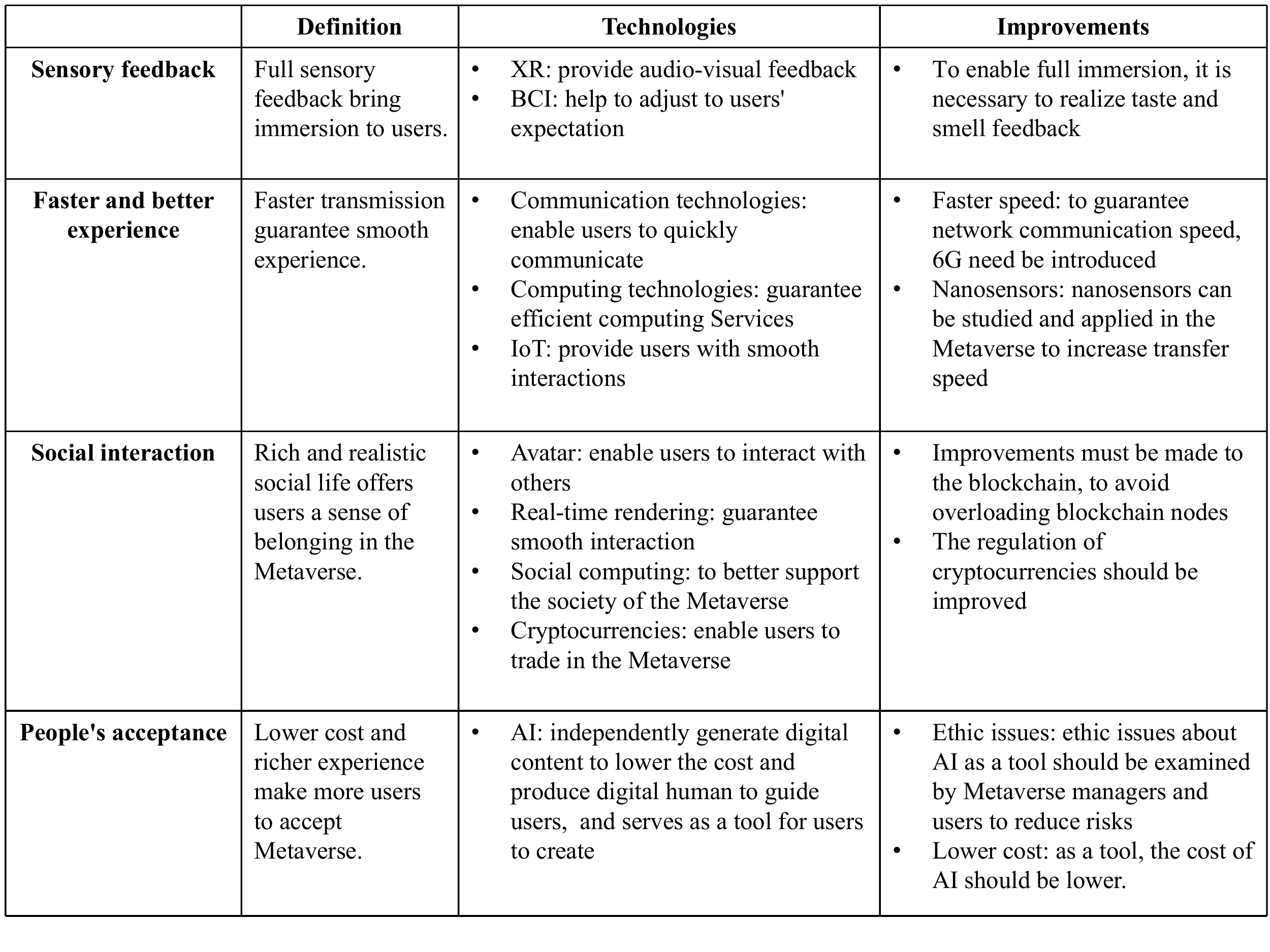}
    \caption{Embodiment of human-centric in technologies.}
    \label{fig:table}
\end{figure}

\end{document}